\documentclass[10pt]{article}
\usepackage{amsmath}
\usepackage{helvet}
\usepackage[a4paper, margin=2cm]{geometry}
\usepackage[english]{babel}
\usepackage{graphicx}
\usepackage{float}
\usepackage{etoolbox}
\usepackage{wasysym}
\usepackage{gensymb}
\usepackage{url}
\usepackage{cite}
\usepackage[utf8]{inputenc}
\usepackage[T1]{fontenc}

\patchcmd{\thebibliography}{\section*{\refname}}{}{}{}

\DeclareMathOperator*{\argmin}{arg min}

\usepackage{wrapfig}
\usepackage[table,xcdraw]{xcolor}

\usepackage{siunitx}
\usepackage{titlesec}

\usepackage{titlesec}

\titleformat{\paragraph}
{\normalfont\normalsize\bfseries}{\theparagraph}{1em}{}
\titlespacing*{\paragraph}
{0pt}{3.25ex plus 1ex minus .2ex}{1.5ex plus .2ex}

\date{\vspace{-10ex}}

\begin{document}

\title{Dynamic laboratory X-ray phase-contrast microtomography with structure-based prior regularisation \\
}
\maketitle

\begin{center}
    \textbf{{\large }}
    \\
    {\large\textbf{Harry Allan}$^{1,2}$}, 
    \large{Tom Partridge}$^{1}$,
    \large{Joseph Jacob}$^{3,4}$,
    \large{Marco Endrizzi}$^{1,2}$
    \\
    \textit{$^1$Department of Medical Physics and Biomedical Engineering, University College London, London, UK} \\

    \textit{$^2$X-ray Microscopy and Tomography Laboratory, The Francis Crick Institute, London, UK} \\

    \textit{$^3$Hawkes Institute, University College London, London, UK} \\

    \textit{$^4$UCL Respiratory, University College London, London, UK} \\

\end{center}

\begin{abstract}

X-ray microtomography is a versatile tool allowing the measurement of the 3D structure of optically thick samples. As a non-destructive technique, it is readily adapted to 4D imaging, where a sample can be monitored over time, and especially in conjunction with the application of external stimuli. To apply this technique with the limited X-ray flux available at a conventional laboratory source, we leverage the contrast enhancement of free-space propagation phase-contrast imaging, achieving an increase in contrast-to-noise ratio of 5.8x. Furthermore, we combine this with iterative reconstruction, using regularisation by a structure-based prior from a high-quality reference scan of the object. This combination of phase-contrast imaging and iterative reconstruction leads to a 29.2x improvement in contrast-to-noise ratio compared to the conventional reconstruction. This enables fully dynamic X-ray microtomography, with a temporal resolution of 9 s at a voxel size of 10.5 \textmu{}m. We use this to measure the movement of a waterfront in the fine vessels of a wooden skewer, as a representative example of dynamic system evolving on the scale of tens of seconds.
\end{abstract}

\section{Introduction}

The penetrating power of hard X-rays makes them a versatile tool for imaging the structure of non-transparent materials across a range of length scales. In combination with tomography, this enables the reconstruction of 3D volumes and the recovery of complex internal sample morphology \cite{withers2021x}. Being a non-destructive technique, X-ray tomography is readily extended to 4D (3 spatial + temporal dimension) imaging. This capability enables longitudinal in-situ testing of the same sample as it evolves with time, often examining the response of the test object to an external stimulus. Rather than acquiring separate 3D snapshots at intermediate time points, high temporal resolution systems allow fully dynamic imaging, in which data is continuously acquired while the test object evolves \cite{marone2020time}. A range of fields have benefitted from in-situ and dynamic X-ray tomography, including food science \cite{babin2006fast,mokso2011following}, additive manufacturing \cite{thompson2016x,du2018x}, and biomedical sciences \cite{murrie2020real,fardin2021imaging,walker2014vivo}.
\par
Achieving sufficient contrast-to-noise ratio (CNR) for tomographic reconstruction and subsequent analysis typically requires high X-ray flux to reduce scan times and enable the capture of fast dynamic processes. Specially designed synchrotron radiation facilities (SRFs) generate extremely bright X-ray beams by the acceleration of electron bunches. In addition to high X-ray flux, SRFs also deliver a coherent beam, allowing users to benefit from phase effects \cite{endrizzi2018x}, rather than just X-ray attenuation. This enables the imaging of low-attenuation objects, and increases CNR compared to the equivalent attenuation only image \cite{kitchen2017ct}. These factors enable the highest temporal resolution at SRFs, with tomography acquisition times down to a few milliseconds \cite{mokso2017gigafrost,yashiro2017sub,garcia2019using}.
\par
Despite a growing number of facilities worldwide, the need for regular longitudinal access makes the development of dynamic X-ray tomography using laboratory X-ray sources an attractive prospect \cite{zwanenburg2021review}. With peak X-ray brilliances (a measurement of the coherent flux) typically 10+ orders of magnitude lower than those at SRFs, it is a challenge to obtain sufficient CNR for valuable analyses. Nevertheless, a number of demonstrations with temporal resolution on the order of 10 s \cite{eggert2014high,dewanckele2020innovations,bultreys2016fast}, and even sub second \cite{vavvrik2017laboratory,makiharju2022tomographic,allan2024sub,parker2024lab} have been demonstrated. Similar to SRFs, XPCI can also be employed at laboratory sources. A range of techniques have been developed to allow XPCI with their lower coherent flux by using optical elements in the beam \cite{pfeiffer2006phase,olivo2007coded,vittoria2015beam,zanette2014speckle}. These methods often provide the possibility for multi-contrast imaging, which can give unique insights into the sample, with dynamic applications in monitoring tissue freezing protocols \cite{john2024x} and in additive manufacturing \cite{massimi2021dynamic}. However, these techniques often require multiple movements of optical elements, which complicates fast and dynamic imaging. Some techniques, particularly beam-tracking \cite{vittoria2015beam} and speckle-based imaging \cite{zanette2014speckle}, can be set up to obtain multi-contrast (including phase) images with a 'single-shot'. However, the inclusion of optical elements still results in a decrease in X-ray flux, and single-shot modes typically also require a sacrifice in spatial sampling (and thus resolution). 
\par
It is for this reason that dynamic XPCI based on free-space propagation (FSP) \cite{wilkins1996phase} is particularly promising for lower brilliance sources. With this optics-less approach, the CNR boost of phase-contrast can be achieved with no loss of flux compared to the purely attenuation-based counterpart. It is interesting to note that the CNR improvement associated with XPCI is not constant with dose \cite{kitchen2017ct}. The CNR boost is more pronounced for shorter exposures, particularly with tomography, making XPCI an incredibly valuable tool for fast dynamic tomography. Outside of large-scale facilities, time-resolved FSP XPCI has been utilised for dynamic XPCI radiography of mouse lung motion using an inverse Compton scattering source \cite{gradl2018vivo}. Extending to 4D, regional lung function of a cystic fibrosis mouse model has been measured using a liquid metal-jet source with a respiratory-gated acquisition \cite{murrie2020real}.
\par
A key development in the enabling of fast and dynamic tomography is the advancement of reconstruction algorithms. Compared to analytical methods, iterative algorithms \cite{beister2012iterative} such as the simultaneous iterative reconstruction technique (SIRT) \cite{gilbert1972iterative} perform well with the undersampled and noisy projection data that are typical of dynamic imaging. Self-supervised deep learning methods such as noise2inverse \cite{hendriksen2020noise2inverse} also offer promising results for handling noisy data, yet suffer when applied to undersampled data. Algorithms that utilise spatio-temporal \cite{ritschl2012iterative} or structure-based \cite{papoutsellis2021core} priors are particularly capable of generating useful reconstructions for time-resolved experiments, by exploiting mutual information between the target reconstruction and some other prior data.
\par
We present dynamic X-ray microtomography with a laboratory source, achieving a temporal resolution of 9 s at a voxel size of 10.5 \textmu{}m. We demonstrate this through the measurement of water uptake in a birch wood skewer, proving quantitatively the CNR gain from applying XPCI to this low-attenuation material. Furthermore, to cope with the relatively low brilliance of the laboratory source, we apply a reconstruction method utilising a structure-based prior, to enable quality reconstruction from fast and undersampled datasets.

\section{Methods}

\subsection{Free-space propagation X-ray phase-contrast imaging}

For X-rays passing through a thin, weakly refracting object, the near-field image intensity recorded by a detector downstream of the object is well described by Fresnel diffraction \cite{paganin2006coherent}. Thus, the intensity at the detector plane $z = z_d$ can be approximated by the transport of intensity equation (TIE) as \cite{wilkins1996phase}

\begin{equation}
    I(x,y,z=z_d) = \frac{I(x,y,z=z_o)}{M^2} \left(
    1 + \frac{R_2 \lambda}{2 \pi M} \nabla_\bot^2 \phi (x,y;\lambda)
    \right),
    \label{fsp_equa}
\end{equation}

where $R_2$ is the propagation distance from the object plane $z = z_o$ to the detector plane, $M = (\text{R}_1 + \text{R}_2) / \text{R}_1$ is the geometric magnification of the image, $\lambda$ is the wavelength of the incident radiation, and $\nabla_\bot^2 \phi (x,y;\lambda)$ is the transverse Laplacian of the object-induced phase shift projected onto the $xy$-plane. It can therefore be shown that for R$_2 = 0$, the image recorded is simply the pure attenuation contact image $I(x,y,z=z_o) = I(x,y,z=0) \, \text{exp} \left( -\int_{o} \mu (x,y,z;\lambda) \, dz \right)$, where $I(x,y,z=0)$ is the X-ray intensity in the absence of the object, and $\int_{o} \mu (x,y,z;\lambda) \, dz$ is the integral along the beam path of the 3D distribution of the linear attenuation coefficient of the material. For a polychromatic X-ray source, as is typically used in laboratory imaging, $I(x,y,z=z_d)$ is instead replaced by the weighted integral across $\lambda$. As the distance R$_2$ is extended, the Laplacian term becomes significant and the image intensity becomes dependent not only on the amplitude of the wave at the object exit plane, but also on its phase. While in the near-field, the phase term grows linearly with increasing $R_2$. However, for finite X-ray source size and detector point spread function (PSF), the resultant measured image is given by the convolution of the total PSF with the input image intensity, thus the fringe contrast depends also on the spatial resolution of the system.
\par
Assuming Gaussian PSFs, for a divergent X-ray geometry with a non-negligible X-ray source size, the spatial resolution of the imaging system can be described by a PSF with width $\sigma$ as

\begin{equation}
    \sigma = \sqrt{
    \left( 1 - \frac{1}{M}   \right)^2 \sigma_s^2
    + \frac{\sigma_d^2}{M^2}
    },
    \label{eq:spatialres}
\end{equation}

where $\sigma_s$ and $\sigma_d$ are the width of the source and the detector PSFs respectively. Thus it can be shown that for a fixed system length $R_1 + R_2$, there exists an optimum magnification $M_\text{opt}$ for which phase fringe contrast is maximised, by balancing the amplification of the Laplacian term and the system spatial resolution. This optimum is dependent only on the ratio of the source and detector PSFs \cite{bidola2015optimization,lioliou2024framework}, and is given by

\begin{equation}
    M_\text{opt} = 1 + \frac{\sigma_d}{\sigma_s}.
    \label{mopt_eq}
\end{equation}

The dependence of the measured intensity on the system resolution makes it necessary to employ X-ray sources with some combination of small focal spots \cite{eckermann2020phase} or high-resolution detectors \cite{esposito2025laboratory} in order to make the phase fringes detectable. 
\par
Under the assumption of a monochromatic, paraxial, scalar X-ray field incident upon a homogenous object, the induced phase-shift may be retrieved from the mixed image as

\begin{equation}
    \phi(x,y) = - \frac{\lambda}{4\pi}\frac{\delta}{\beta} \, \text{log} \left(\,    
    \left|\mathcal{F}^{-1}\left(\frac{\mathcal{F}(I(x,y,z=z_d)/I(x,y,z=0))}{\frac{ \lambda R_2}{4 \pi M}\frac{\delta}{\beta}(u^2+v^2) + 1}\right)\right| 
    \,\right),
    \label{paganin_eq}
\end{equation}

where $\frac{\delta}{\beta}$ is the ratio of the real to imaginary parts of the object's complex refractive index, and $(u,v)$ are the spatial frequency coordinates corresponding to the real space $(x,y)$ detector pixel coordinates \cite{paganin2002simultaneous}. Violating the stated assumptions by replacing $\lambda$ with the mean polychromatic wavelength $\lambda_\text{eff}$, and applying the technique to real non-homogenous objects, a high-contrast non-quantitative approximation of the phase-shift is retrieved. 
\par

\subsection{Tomographic reconstruction}

The aim of tomographic reconstruction is to solve the linear inverse problem

\begin{equation}
    Au = b,
\end{equation}

where $A$ is the X-ray forward projection operator, $u$ is the volume to be recovered, and $b$ is the measurement data \cite{jorgensen2021core}. For well-sampled and low-noise measurement data, analytical reconstruction based on methods such as the Feldkamp-Davis-Kress (FDK) algorithm \cite{feldkamp1984practical} yields satisfactory reconstructions of $u$. For the reconstruction of noisy or undersampled data, analytical methods may not be sufficient, thus iterative methods become powerful alternatives. 
\par
In the case of dynamic X-ray tomography, it may be possible to obtain a high-quality reference scan of the sample before the initiation of the dynamic process. If the sample remains structurally similar to its initial state throughout the duration of the dynamic imaging, then the reconstruction $v$ of the high quality reference can be used as a structure-based prior for regularisation of the iterative reconstruction \cite{chen2008prior,myers2011dynamic,kazantsev2014novel,papoutsellis2021core}.
\par
\begin{figure}[]
    \centering
    \includegraphics[width=1\linewidth]{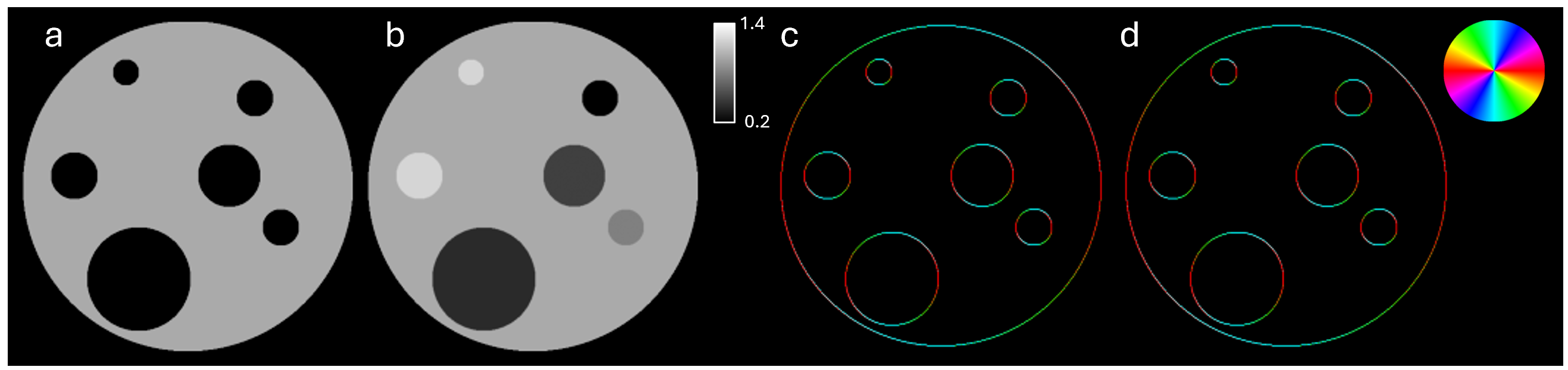}
    \caption{Numerical example showing a slice through an initial volume (a) that transforms through some dynamic process into a new state (b). The corresponding vectors describing the absolute direction of the image gradients in a and b are illustrated in c and d respectively. Because of the temporal evolution, the slices through the volume in (a) and (b) appear different, however the mutual information encoded in the normalised gradient directions remains constant.}
    \label{fig:grad_Plots}
\end{figure}

We justify this by considering some initial volume that transforms through some dynamic process into a new state, illustrated by the slices shown in figure \ref{fig:grad_Plots}a and \ref{fig:grad_Plots}b. This could be representative of the filling of structures with a fluid, or changes in density of a structure. Despite containing some mutual information, there is no longer a 1:1 correspondence between the two states. To capture the existing mutual information, we examine the gradients of the two images. The resultant normalised direction of image gradients effectively captures this information, and is equal for both images, under the assumption that the shape of the structures remains constant. We quantify this for the illustrated example by the change in Pearson's correlation coefficient of 0.886 between the two images, versus 1.000 for their gradient directions.

It has been shown that this vectorial information can be harnessed to regularise certain inverse problems, by capturing this mutual information \cite{ehrhardt2013vector}. To apply this to tomography, we aim to solve the inverse problem

\begin{equation}
    u(t) = \argmin_{u(t)\geq0} \frac{1}{2} || Au(t) - b ||^2 + \alpha \, d\text{TV}(u(t),v), 
    \label{minimisation_eq}
\end{equation}

where $u(t)\geq0$ enforces non-negativity of the solution of $u$ at time $t$, $\alpha$ controls the strength of the regularisation, and $d\text{TV}(u(t),v)$ is the directional total variation regulariser, defined as \cite{papoutsellis2021core}

\begin{equation}
    d\text{TV} := ||D_v\nabla u(t) ||_{2,1} = \sum_{x,y,z}(|| D_v\nabla u(t) ||_2)_{x,y,z}.
\end{equation}

The weight matrix $D_v$ captures the orthogonal components of the normalised gradient field $\nabla v$, incorporating a Tikhnonov regularisation parameter $\eta$ to ensure numerical stability. Thus $d$TV regularisation encourages the solution to align with the edges and features present in the reference image $v$. To separately handle the smooth data fidelity term and the non-smooth $d$TV term, the inverse problem can be efficiently solved using the primal-dual hybrid gradient (PDHG) method.

\subsection{Experiments}

The experiment was carried out at the NXCT (National research facility for lab-based X-ray Computed Tomography) multi-contrast X-ray micro-CT system \cite{i2024new}, utilising a Rigaku MicroMax 007-HF rotating molybdenum anode X-ray source. The source was operated at 50 kV tube voltage and 24 mA current. A custom detector based on a scintillator and lens-coupled sCMOS camera with a pixel size of \SI{15}{\micro\meter} was placed 460 mm away from the source. For a source of width \SI{70}{\micro\meter} and a detector with line spread function (LSF) of \SI{30}{\micro\meter}, $M_\text{opt}$ according to equation \ref{mopt_eq} is 1.43 and thus $R_2$ was set to 140 mm. This geometry resulted in an effective pixel size at the sample plane of \SI{10.5}{\micro\meter}.
\par
Water uptake in a thin birch wood skewer was used as a dynamic process to test the system. This process involves the filling of vessels in the wood via capillary action, in analogy to the uptake of water and nutrients in living plants. This process does not impact the overall structure or shape of the sample, thus making it a good candidate for reconstruction using a structure-based prior. 
\par
A high-quality reference scan was acquired using 1001 projections of 1 s equally distributed through 360$\degree$. Dark and flat frame corrected projections were retrieved using the TIE-based equation \ref{paganin_eq}. Retrieval was carried out using $\lambda$ corresponding to a mean energy of 19 keV, with $\frac{\delta}{\beta} = 1000$ chosen to optimise CNR and spatial resolution by visual inspection. The high-quality reference volume $v$ was reconstructed analytically using the FDK algorithm implemented in the Core Imaging Library (CIL) \cite{jorgensen2021core}.
\par
The dynamic scans were acquired under continuous sample rotation at an angular velocity of 20$\degree$/s, with the detector operating in continuous acquisition mode with an exposure time of 50 ms. Projections were retrieved using the same parameters as for the high-quality reference scan. Reconstruction from each set of 180 projections results in a temporal resolution of 9 s per full CT scan. 
\par
To effectively utilise the $d$TV regularisation, it is vital that the reference scan $v$ is well aligned with the dynamic data. As the dynamic scan was operated in a flyscan mode, it was necessary to find the angular offset between the start angle of the two scans to ensure alignment of the reconstructions. This was accomplished by minimising the $\mathit{l}^2$ norm between $v$ and $u(t)$, across some range of angular offsets (see supplementary material \ref{sup:offset_min}). A number of frames (16) were skipped by the detector around the midway point of the scan, leading to a change in the offset. Thus the alignment step was taken for each $u(t)$ to ensure consistent alignment with $v$ across all $t$. A similar process was also used to estimate the correct angular step between projections, to account for the finite detector readout and dead time (see supplementary material \ref{sup:step_min}).
\par
Reconstruction of the dynamic data was accomplished by solving equation \ref{minimisation_eq} using the PDHG method in CIL \cite{jorgensen2021core,papoutsellis2021core} for each time step $t$. For each $t$, $u(t)$ was initialised with an FDK reconstruction $u_\text{FDK}(t)$ to speed convergence, followed by 10 PDHG iterations with $\alpha = 5 \times 10^{-9}$ and edge parameter $\eta = 5 \times 10^{-11}$, both chosen by visual inspection to optimise the CNR. We will present reconstructions acquired by direct analytic reconstruction of attenuation data, direct analytic reconstruction of phase-retrieved data, and $d$TV regularised reconstruction of phase-retrieved data, which will be henceforth referred to as conventional, phase-retrieved, and regularised phase-retreived respectively.

\section{Results and discussion}

\subsection{Demonstration of phase-contrast edge enhancement}

\begin{figure}[]
    \centering
    \includegraphics[width=1\linewidth]{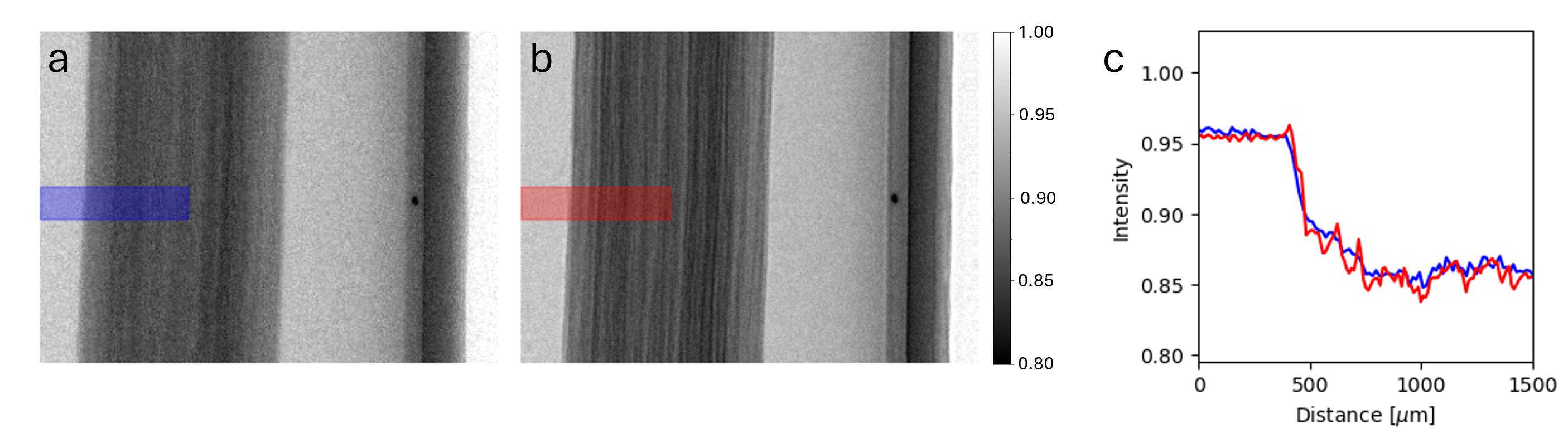}
    \caption{Comparison of the same birch wood skewer imaged with a propagation distance of $R_2 \approx$ 0 mm (a) and $R_2 =$ 140 mm (b). Plotted line profiles illustrate increased structure and phase fringes at the sample edges. Both images are the average of 5 x 3 s exposures, indicating an increase in sample contrast at the same exposure.}
    \label{fig:static_scans}
\end{figure}

To demonstrate the potential of FSP XPCI using the lab-based system, the same birch skewer sample was imaged with $R_2$ = 140 mm, and compared to the equivalent contact image with $R_2 \approx$ 0 mm. Much more structure is visible in the wood in the FSP image in figure \ref{fig:static_scans}b compared to the contact image in figure \ref{fig:static_scans}a. The line profiles plotted in figure \ref{fig:static_scans}c show clear edge enhancement in regions of the skewer that appear relatively homogenous in the contact image. It is important to note that both images were obtained from the average of 5 x 3 s exposures acquired with the same source-to-detector distance, with the only difference being the propagation distance $R_2$. Phase effects and edge enhancement were therefore obtained without any loss of flux or change in the acquisition process, thus making FSP XPCI an efficient method for dynamic imaging.

\subsection{Effect of phase-retrieval and regularised reconstruction}

\begin{figure}[]
    \centering
    \includegraphics[width=1.0\linewidth]{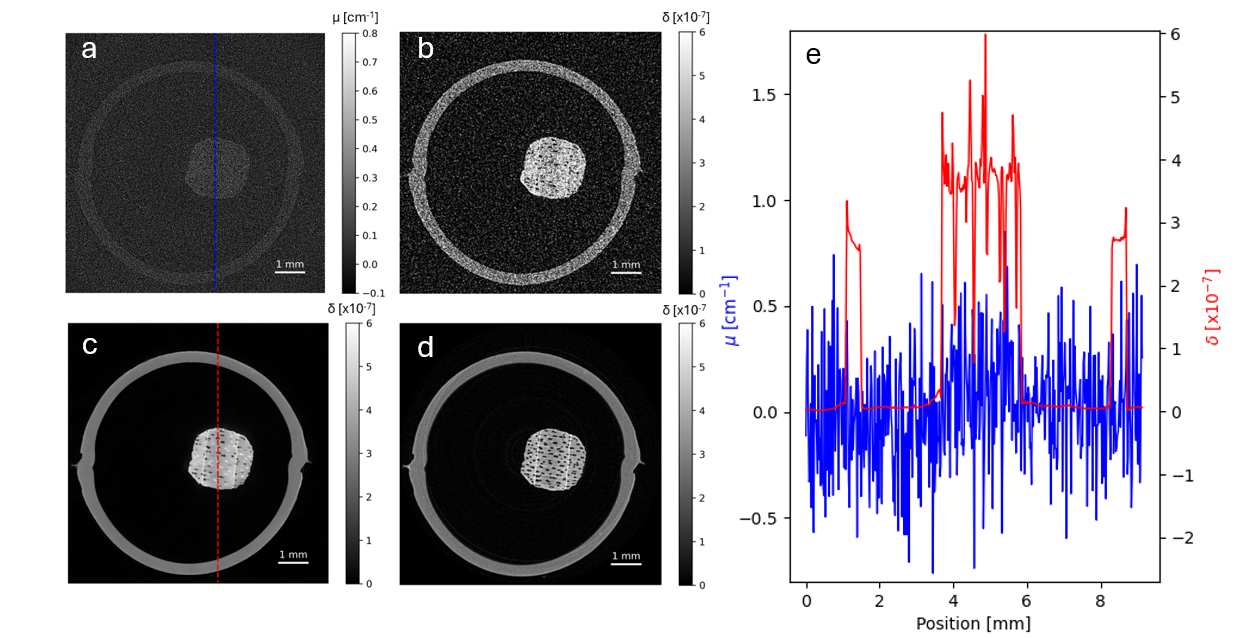}
    \caption{Axial slices of the reconstructed wooden skewer during the 9 s temporal resolution dynamic scan, conventional (a), phase-retrieved (b), and regularised phase-retrieved (c). Axial slice of the high-quality reference reconstruction, which was used as a structure-based prior for the regularisation (d). Line profiles (e) from the slices in (a) and (c) demonstrate the much greater CNR of the regularised phase-retrieved reconstruction, despite using the same raw data as the conventional reconstruction.}
    \label{fig:slices_and_profiles}
\end{figure}

Figure \ref{fig:slices_and_profiles} shows the same axial slice of the conventional (\ref{fig:slices_and_profiles}a), phase-retrieved (\ref{fig:slices_and_profiles}b), and regularised phase-retrieved (\ref{fig:slices_and_profiles}c) reconstructions. All slices were reconstructed using the same 9 s of input data, immediately after beginning dynamic micro-CT acquisition. Figure \ref{fig:slices_and_profiles}d shows the same axial slice of the same sample in a dry state, before initiating the dynamic experiment. This was reconstructed from $\sim$16 minutes of well sampled data, providing a high-quality reconstruction to act as a structure-based prior for $d$TV regularisation. Line profiles plotted in figure \ref{fig:slices_and_profiles}e corresponding to the conventional and the regularised phase-retrieved reconstruction illustrate the much superior CNR achieved using the same dataset with appropriate data processing and advanced reconstruction methods. In figure \ref{fig:slices_and_profiles}a, only the approximate shape of the sample can be made out, with no structure visible. In figure \ref{fig:slices_and_profiles}b, after phase-retrieval, the structure of the vessels start to appear but are largely obscured by noise. Finally, in figure \ref{fig:slices_and_profiles}c, the air filled vessels are well delineated from the wood, and the water filled vessels now become visible.

\begin{figure}[]
    \centering
    \includegraphics[width=0.97\linewidth]{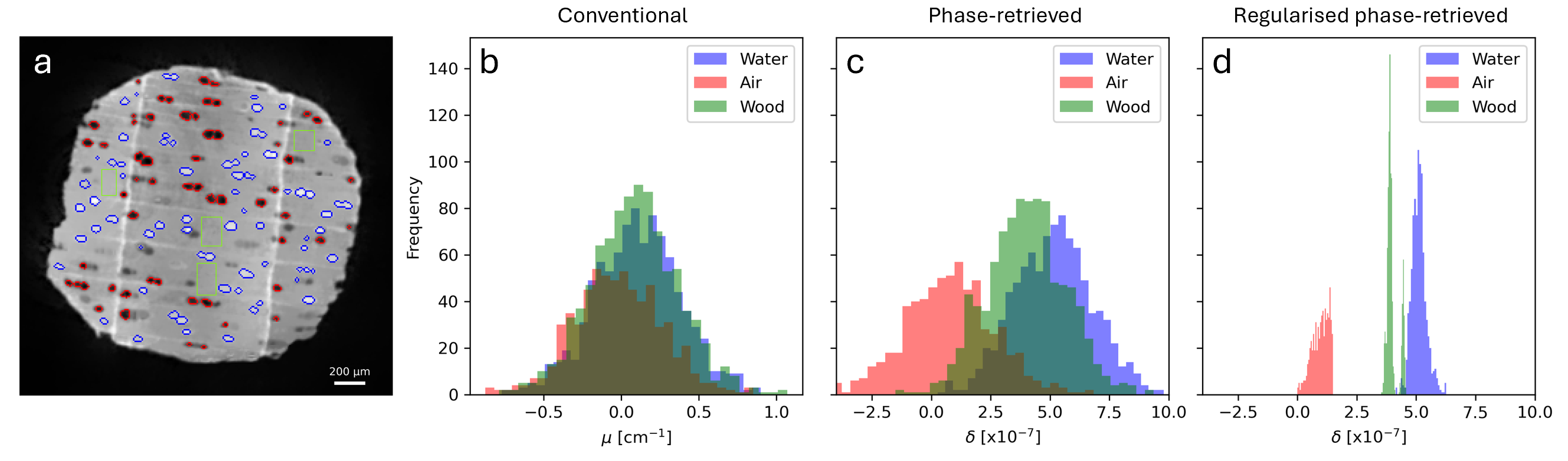}
    \caption{Segmented axial slice of the regularised phase-retrieved reconstruction (a). Red-bordered regions indicate unfilled vessels, while blue-bordered regions indicate water filled vessels. The green rectangles indicate regions of the wood used to measure the distribution of grey-values. Histograms of voxels falling into the water, air, or wood categories are shown for the conventional (b), the phase-retrieved (c), and the regularised phase-retrieved reconstruction (d).}
    \label{fig:segment_and_histos}
\end{figure}

To demonstrate quantitatively the practical impact of the phase-retrieval and regularised reconstruction, figure \ref{fig:segment_and_histos}a shows an axial slice complete with an overlaid segmentation of air-filled and water-filled vessels. Some of the vessels feature intermediate grey-values (and are thus excluded from the following analysis), indicating that the waterfront reached this vertical position during the finite acquisition time of this time point. Using the segmented vessel voxels and the representative wood regions indicated on figure \ref{fig:segment_and_histos}a by the green rectangles, histograms of each category were computed for each of the reconstructed slices. With each additional processing step, moving from figure \ref{fig:segment_and_histos}b to figure \ref{fig:segment_and_histos}d, it is seen that the grey-values belonging to each category become more separated. Having initially segmented vessels using the high-quality reference volume, it is important for further analysis that voxels within the vessels can subsequently be accurately categorised as water or air. The mean values and standard deviations of voxels categorised as air, birch, and water are tabulated in table \ref{tab:vals}. We note that both the analytical and regularised reconstructions of the phase-retrieved data yield comparable grey-values, indicating that the regularisation does not affect the quantitativeness of the reconstruction. This is particularly insightful for the reconstruction of the water, which retains comparable grey-value despite being regularised by a reference which did not contain water. To enable a direct comparison of each method, the CNR of water compared to air was calculated, where CNR = $(\Bar{\text{water}} - \Bar{\text{air}}$)/$\sigma$, with $\sigma$ defined as the mean standard deviation of the two categories. The resultant CNR$_\text{conv}$ = 0.43, CNR$_\text{PR}$ = 2.50, and CNR$_\text{PRreg}$ = 12.55 indicate the much improved category separation after phase-retrieval (a factor of 5.8x) and after regularised reconstruction (a factor of 29.2x).

\begin{table}[h]
    \centering
    \renewcommand{\arraystretch}{1.3}
    \begin{tabular}{lccc}
        \hline
        \textbf{Material} & \textbf{Conventional} [cm$^{-1}$] & \textbf{Phase-retrieved} [$\times 10^{-7}$] & \textbf{Regularised phase-retrieved} [$\times 10^{-7}$]
        \\
        \hline
        Air   & $-0.02 \pm 0.28$ & $0.81 \pm 1.85$ & $0.97 \pm 0.35$ \\
        Birch & $0.08 \pm 0.28$  & $4.00 \pm 1.65$  & $3.98 \pm 0.24 $ \\
        Water & $0.10 \pm 0.28$  & $5.14 \pm 1.62 $ & $5.11 \pm 0.31$ \\
        \hline
    \end{tabular}
    \caption{Comparison of conventional, phase-retrieved, and regularised phase-retrieved grey-values (mean $\pm$ standard deviation) for each material category.}
    \label{tab:vals}
\end{table}

\subsection{Spatial resolution}

\begin{figure}[]
    \centering
    \includegraphics[width=0.99\linewidth]{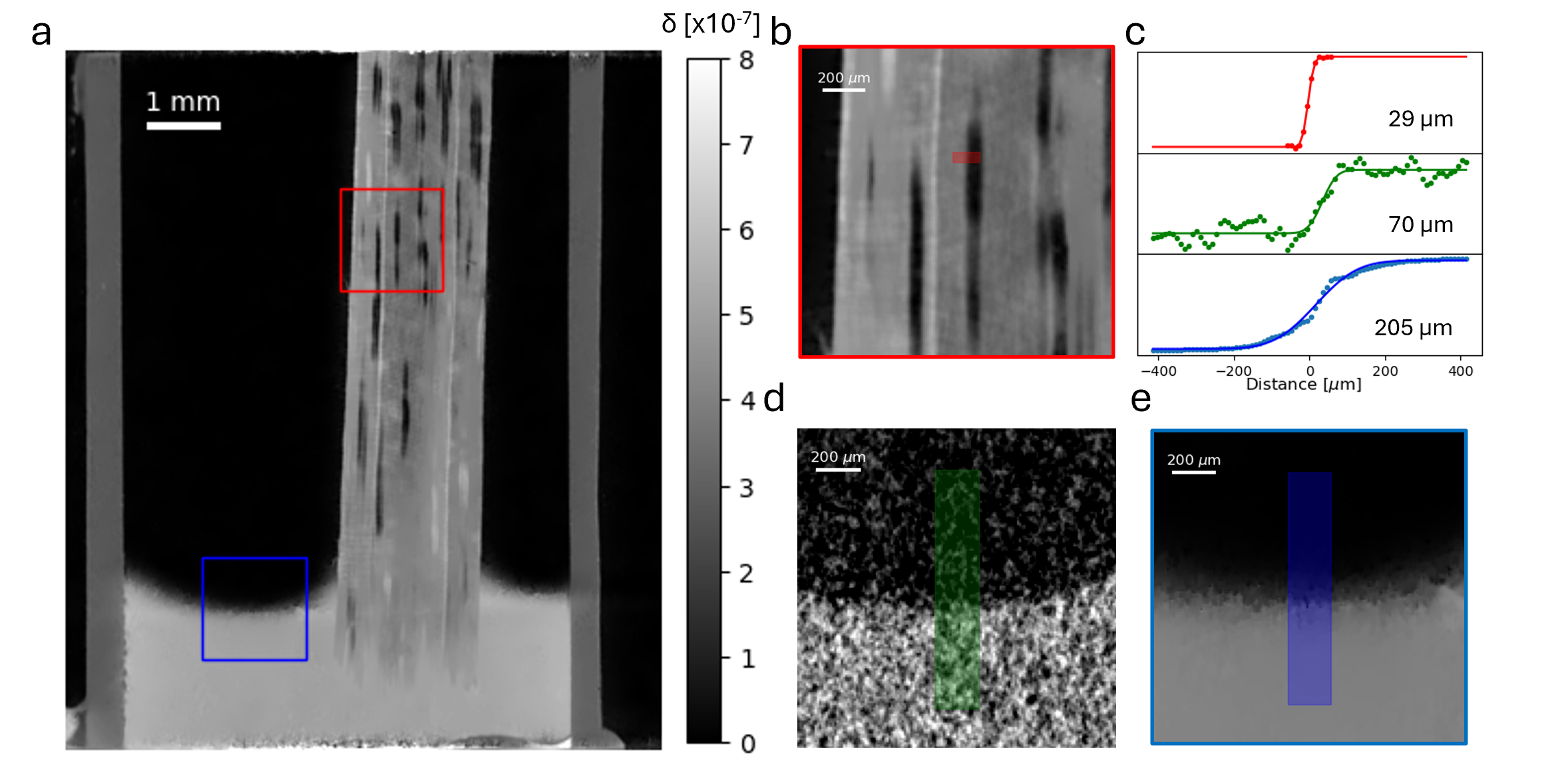}
    \caption{A coronal slice of the regularised phase-retrieved volume $u(t = 9 s)$. Zoomed regions are indicated on the figure, illustrating structure within the skewer (b), and the water level within the container (e). Edge response functions are taken (c) to characterise the spatial resolution in a static region of the regularised phase-retrieved reconstruction (b) in red, a moving region of the same reconstruction (e) in blue, and the equivalent moving region in the conventional reconstruction (d) in green. The highest spatial resolution is observed in the static case where it matched the theoretical resolution of the imaging system. The regularisation vastly improves on the SNR of dynamic regions, however with a threefold compromise on the achieved spatial resolution.}
    \label{fig:spatial_res}
\end{figure}

Figure \ref{fig:spatial_res}a shows a coronal slice of the volume $u(t = 9 s)$ reconstructed using the regularised phase-retrieval method. To characterise the spatial resolution of the reconstructions, edge response functions (ERFs) were taken at a number of locations. From each of these, the Gaussian equivalent full-width-at-half-maximums (FWHMs) were calculated. The ERF taken at a vessel edge in figure \ref{fig:spatial_res}b represents a static region of the sample, in which no water transport has yet occurred, and thus the gradients of $v$ and $u(t = 9 s)$ are very well aligned. This yielded a FWHM of 29 \textmu{}m. In comparison, the same region of the conventional dynamic reconstruction yielded a FWHM of 31 \textmu{}m. Both are consistent with the 30 \textmu{}m estimated using equation \ref{eq:spatialres}.

We show also the spatial resolution at the water surface in figure \ref{fig:spatial_res}e. We note that in this region the sample is affected by additional blur due to the finite temporal resolution of the system. As the reference scan was taken of the completely dry sample before any water inclusion, the gradients of $v$ and $u(t)$ are not aligned. As the minimisation procedure encourages the alignment of the gradients with each other, the lack of a gradient in $v$ causes the algorithm to penalise any gradient appearing in $u(t)$. The resultant FWHM of 205 \textmu{}m is thus worse than the 70 \textmu{}m corresponding to the conventional reconstruction, shown in figure \ref{fig:spatial_res}d. While being a potential limitation of the method, we emphasise that this impacts only this region of the reconstruction in which the reference and target gradients are not parallel. Even with the inclusion of water in the dynamic scan, vessels are still accurately reconstructed due to the presence of gradients in the reference volume.

\subsection{Capillary uptake of water in birch vessels}

\begin{figure}[]
    \centering
    \includegraphics[width=1\linewidth]{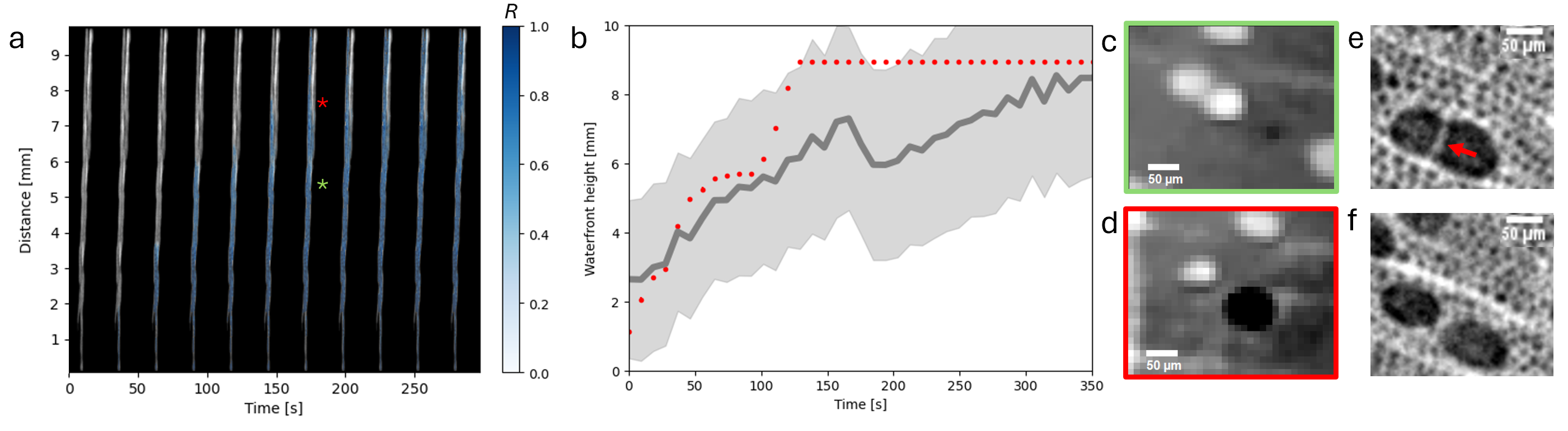}
    \caption{The progression of water up a single vessel of the skewer is visualised throughout the duration of the scan (a). The ratio $R$ of water thickness to total vessel thickness is overlaid in blue at intervals of 27 s. The progression of the waterfront in the vessel shown in (a) is plotted with red markers in (b), error bars are included but are smaller than the markers. In dark grey is the median waterfront height, while the light grey filled region indicates the standard deviation across all vessels. A sharp jump in the waterfront height at around 100 s corresponds to the branching from joined vessels (c), into separated vessels (d). The approximate locations of (c) and (d) are indicated by the colour matched asterisks on (a). A higher resolution scan of a similar sample shows the structure in greater detail, where two vessels are joined by a thin membrane (e), but later separate to become independent (f).}
    \label{fig:waterfronts_plot}
\end{figure}

The progression of the waterfront up the skewer was characterised for each vessel independently. Voxels within the vessels were classified as containing water when the difference in voxel grey-value $\delta_{u(t)}$ - $\delta_{v}$ was greater than the halfway point  between the classes, $(\Bar{\delta}_\text{water} - \Bar{\delta}_\text{air})/2 = 2.07 \times 10^{-7}$. Following this, $R$ was calculated as the ratio of the projected thickness in the coronal plane of water containing voxels to the projected thickness of the vessel. The progression of the waterfront up the skewer is visualised for a single vessel in figure \ref{fig:waterfronts_plot}a. 

The waterfront height in a given vessel was subsequently extracted by finding the height at which the total ratio of water to vessel thickness drops below 0.1, and remains below this threshold for a distance of 20 pixels (210 \textmu m). The evolution of the waterfront height in the vessel in figure \ref{fig:waterfronts_plot}a is plotted with red markers in figure \ref{fig:waterfronts_plot}b. Note that error bars were calculated by varying the ratio threshold between 0.05 and 0.15, but are smaller than the markers and thus are not visible. The dark grey line on the same plot indicates the median waterfront height across all vessels, while the light grey filled region indicates the standard deviation across all vessels. A large part of the variance in waterfront height across vessels comes from differences in the initiation time of water entering a given vessel, as well as branching structures causing an uneven flow rate for some vessels.
\par
In particular, the individually plotted vessel shows a sharp jump in waterfront height at around 100 s. Closer examination in figure \ref{fig:waterfronts_plot}c shows that this corresponds to the water flow shifting from movement through a joined vessel pair, which then separates, with the water only continuing to move in a single vessel, illustrated in figure \ref{fig:waterfronts_plot}d. Figure \ref{fig:waterfronts_plot}e and \ref{fig:waterfronts_plot}f illustrate this structure in a higher resolution scan (see supplementary material \ref{sup:hires}) of a similar sample. Initially, two vessels are connected by a thin membrane, through which water may permeate allowing joint movement of the waterfront through a large effective vessel diameter. These vessels then separate, and it is possible that water may continue to flow through only one vessel, taking the path of least resistance. As the capillary rise is inversely proportional to the vessel radius, this results in a rapid advancement of the waterfront height. There are indications that the exact grouping of vessels within a network are an adaption to reduce the risk of embolism \cite{carlquist1984vessel,mrad2018network}, allowing fluids to continue undisrupted through interconnected vessels even in the presence of obstructions. This suggests that our methodology may be suitable for the measurement and monitoring of such processes, providing time-resolved insights at the microscale. 
\par
We applied our system and methodology to the measurement of water flow through a wooden skewer via capillary action, which was chosen as representative of a class of experiments in the study of fluid transport. In particular, the method proposed here could be applied to the measurement of water flow in and around plants using both X-ray \cite{kim2010synchrotron,daly2018quantification} and neutron tomography \cite{warren2013neutron,totzke2017capturing}. Similarly, the measurement of fluid flow through multi-phase materials and pore networks is an active research area for which the structural similarity of image features throughout the course of a scan could be leveraged to improve CNR \cite{bultreys2016fast,menke2015dynamic,andrew2015imaging}. 
\par
Potential clinical uses for the methodology include estimation of blood flow or airflow in animal models of health and disease \cite{jung2012vivo,park2016x,asosingh2022preclinical}. Labelled radiodense particles could for example be used to measure flow perturbation at different severities of tissue damage in various organs of interest. In the pharmacological domain, such methods could also be used to estimate the distribution and mechanics of inhaled substances in the airways following deposition of therapeutic compounds, as has been demonstrated using radiography \cite{gradl2019visualizing,donnelley2013variability}. This could help determine whether active molecules are reaching their desired target location, as well as providing an experimental framework with which to optimise drug delivery using aerosol or nebulised routes of administration.
\par
In applications such as in-vivo biological imaging where larger sample structure change is encountered, a single reference scan would not be sufficient to capture the expected image gradients. In such cases, the demonstrated improvement in CNR associated with phase-retrieval would be an important advantage. Alternatively, future work could investigate other reconstruction methods based on mutual information shared between time-points \cite{ritschl2012iterative}, without utilising a high-quality reference.

\section{Conclusion}

We have presented an experimental setup and data processing methodology for dynamic X-ray phase-contrast microtomography on the seconds timescale with a compact setup. The use of a detector with sufficient spatial resolution to resolve phase fringes, in conjunction with phase-retrieval, allowed substantial (5.8x) improvements in the contrast-to-noise ratio. This was based on free-space propagation of the X-ray beam and did not require additional optical elements, with the benefit of exploiting all of the available flux. The absence of optical elements simplifies the data acquisition by removing potential constraints linked to their alignment or stepping.  Furthermore, we have shown iterative reconstruction using structure-based $d$TV regularisation. The combination of these two techniques generated a notable increase in the contrast to noise ratio of $\sim$29x, allowing accurate segmentation and classification of voxels despite the relatively fast exposures. We demonstrated that in regions where the gradients of the reference and dynamic reconstruction are aligned, the achievable spatial resolution is consistent with theoretical predictions, despite using undersampled data. We envisage a range of potential applications for the demonstrated approach, including, for example plant research, pore-scale flow experiments, and flow estimation in animal models.

\section{Acknowledgements}

This work is supported by the EPSRC-funded UCL Centre for Doctoral Training in Intelligent,
Integrated Imaging in Healthcare (i4health) (EP/S021930/1), the Department of Health’s
NIHR funded Biomedical Research Centre at University College London Hospitals, and the
National Research Facility for Lab X-ray CT (NXCT) through EPSRC grants EP/T02593X/1
and EP/V035932/1; and by the Wellcome Trust 221367/Z/20/Z. This work is also supported by the Francis Crick Institute, which receives its core funding from Cancer Research UK (CC0102), the UK Medical Research Council (CC0102), and the Wellcome Trust (CC0102).

\bibliographystyle{ieeetr}

\section{Supplementary material}

\subsection{Optimisation of reference scan angular offset}

\label{sup:offset_min}

During dynamic acquisition, projections were acquired in flyscan mode, that is continuous image acquisition to the PC RAM, under constant sample rotation. Such a mode reduces the dead-time in the acquisition procedure, as there is no need to wait for motor movements or for image saving. To account for the finite angular acceleration of the rotation stage, the rotation was initiated and allowed to stabilise for several seconds before beginning image acquisition. As a consequence, the exact start angle of the dynamic acquisition differs from that for the reference scan by some angle $\Delta\theta$. It is vital to estimate and account for this offset in the definition of the reconstruction geometry for the dynamic scan, in order to ensure adherence to the assumption that the reference $u(t)$ and dynamic $v$ volumes are structurally similar.
\par
We achieve this by finding the $\Delta\theta$ that minimises the $\mathit{l}^2$ norm

\begin{equation}
    \Delta\theta = \argmin_{\Delta\theta}
    \left( 
    ||u(t,\Delta\theta + \theta)v(\theta) ||_2
    \right),  
\end{equation}

where $\theta$ is the set of angles used to reconstruct the reference scan. Initially, $||u(t,\Delta\theta + \theta)v(\theta) ||_2$ is calculated for a range of choices of $\Delta\theta$, followed by analytical quadratic interpolation of the minimisation landscape around the minimum, to find the optimum $\Delta\theta$.
\par
\begin{figure}[]
    \centering
    \includegraphics[width=0.6\linewidth]{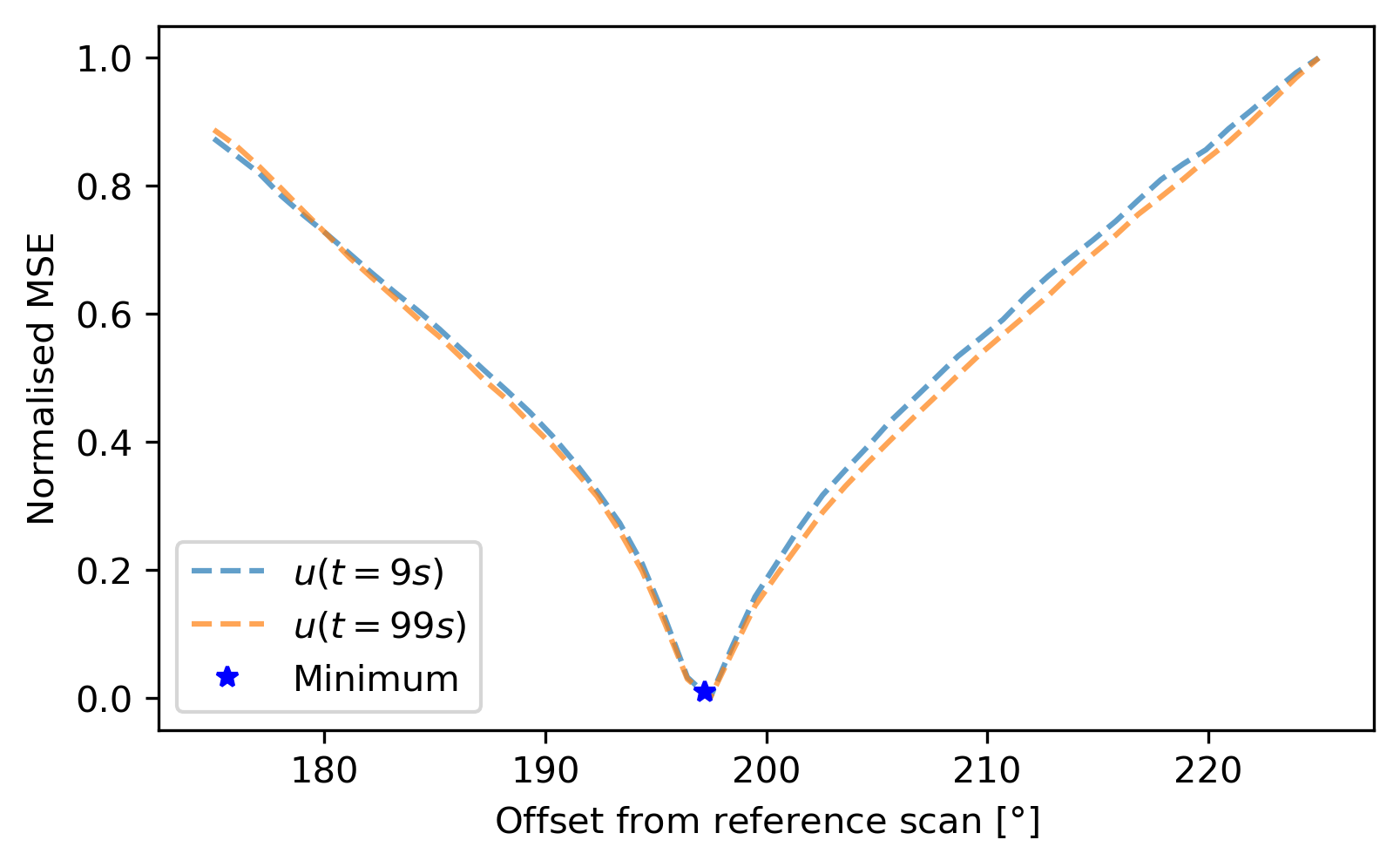}
    \caption{Minimisation landscape of $||u(t,\Delta\theta + \theta)v(\theta) ||_2$ used to find the angular offset $\Delta\theta$ between the start angle of the reference and dynamic scans. The minimum is indicated by the blue star. The same process is repeated for $u(t = 9 s)$ and also $u(t = 99 s)$, demonstrating the stability of the metric even after dynamic evolution of the sample.}
    \label{fig:offset_min}
\end{figure}

Figure \ref{fig:offset_min} illustrates the minimisation landscape resulting from the search for the optimum $\Delta\theta$. To demonstrate the robustness to noise and also the dynamic evolution of the sample, the search was carried out for both $u(t = 9 s)$ and $u(t = 99 s)$, resulting in an almost identical minimum for both cases.

\subsection{Optimisation of dynamic scan angular step}

\label{sup:step_min}

Operating in flyscan mode required that the rotation axis angular velocity and frame rate were matched to acquire the desired number of projections at the correct angular and temporal spacing. The true angular step may deviate from the calculated due to minor differences in experimental readout and dead time. The effect of this mismatch is a volume that appears to rotate throughout the duration of the experiment, rather than remaining stationary. A similar process to that employed in supplementary \ref{sup:offset_min} can be used to correct for this, by finding the angular step $\delta\theta$ that minimises the difference between a dynamic reconstruction $u(a)$ and a dynamic reconstruction at a later time $u(b)$. This can be described by 

\begin{equation}
    \delta\theta = \argmin_{\delta\theta}
    \left( 
    ||u(a,\theta(\delta\theta))u(b,\theta(\delta\theta)) ||_2
    \right),  
\end{equation}

where $\theta(\delta\theta)$ is the set of angles $\theta$ with an angular spacing of $\delta\theta$. We again find the minimum of this function using analytical interpolation.

\subsection{High-resolution microtomography}

\label{sup:hires}

To further investigate the branching structure of the vessel network, a high-resolution micro-CT scan was acquired of a similar birch skewer sample. A detector with a different optical configuration, resulting in an effective pixel size of 3.4 \textmu{}m, was utilised. To account for the higher spatial resolution, the geometrical magnification was adjusted, resulting in an effective voxel size of 3 \textmu{}m. 700 projections of 0.6 s exposure, across a range of 180\degree{}, were acquired in a flyscan mode, followed by single-distance phase-retrieval, and analytical reconstruction using the FDK algorithm in CIL.

\label{hires}

\end{document}